\documentclass{PoS}

\newcommand{\mps}{m_{\rm{PS}}}
\newcommand{\fps}{f_{\rm{PS}}}
\newcommand{\ZP}{Z_{\rm{P}}}

\newcommand{\Oasq}{\mathcal{O}(a^2)}

\title{Light hadrons from $N_f=2+1+1$ dynamical twisted mass fermions}

\ShortTitle{Light hadrons from $N_f=2+1+1$ dynamical twisted mass fermions}

\author{\mbox{R. Baron}$^a$, \mbox{B. Blossier}$^{b}$, \mbox{P. Boucaud}$^{b}$, \mbox{J. Carbonell}$^{c}$, \mbox{A. Deuzeman}$^{d}$, \mbox{V. Drach}$^{e}$, \mbox{F. Farchioni}$^{f}$, \mbox{V. Gimenez}$^{g}$, \mbox{G. Herdoiza}$^{e}$, \mbox{K. Jansen}$^{e}$, \mbox{C. Michael}$^{h}$, \mbox{I. Montvay}$^{i}$, \mbox{E. Pallante}$^{j}$, \mbox{O. P\`ene}$^{b}$, \speaker{S. Reker}\thanks{For the ETM Collaboration} $^{j}$\footnote{Email: {s.f.reker@rug.nl}. Preprint numbers: HU-EP-10/75, SFB/CPP-10-110, MS-TP-10-22, DESY 10-205}, \mbox{C. Urbach}$^{k}$, \mbox{M. Wagner}$^{l}$ and \mbox{U. Wenger}$^{d}$\\
\\\llap{$^a$}CEA, Centre de Saclay, IRFU/Service de Physique Nucl\'eaire, F-91191 Gif-sur-Yvette, France\\
\llap{$^b$}Laboratoire de Physique Th\'eorique (B\^at. 210), Universit\'e de Paris XI, Centre d'Orsay, 91405 Orsay-Cedex, France\\
\llap{$^c$}Laboratoire de Physique Subatomique et Cosmologie, 53 avenue des Martyrs, 38026 Grenoble, France\\
\llap{$^d$}Albert Einstein Center for Fundamental Physics, Institute for Theoretical Physics, University of Bern, Sidlerstr. 5, CH-3012 Bern, Switzerland\\
\llap{$^e$}NIC, DESY, Platanenallee 6, D-15738 Zeuthen, Germany\\ 
\llap{$^f$}Universit\"at M\"unster, Institut f\"ur Theoretische Physik, Wilhelm-Klemm-Stra\ss e 9, D-48149 M\"unster, Germany\\
\llap{$^g$}Dep. de F\'isica Te\`orica and IFIC, Universitat de Val\`encia-CSIC, Dr.Moliner 50, E-46100 Burjassot, Spain\\
\llap{$^h$}Division of Theoretical Physics, University of Liverpool, L69 3BX Liverpool, United Kingdom\\
\llap{$^i$}Deutsches Elektronen-Synchrotron DESY, Notkestr. 85, D-22603 Hamburg, Germany\\
\llap{$^j$}Centre for Theoretical Physics, University of Groningen, Nijenborgh 4, 9747 AG Groningen, the Netherlands\\
\llap{$^k$}Helmholtz-Institut f{\"u}r Strahlen- und Kernphysik (Theorie) and Bethe Center for Theoretical Physics, Universit{\"a}t Bonn, 53115 Bonn, Germany\\
\llap{$^l$}Humboldt-Universit\"at zu Berlin, Institut f\"ur Physik, Newtonstra\ss e 15, D-12489 Berlin, Germany\\}

\abstract{We present results of lattice QCD simulations with mass-degenerate up and down and mass-split strange and charm ($N_f = 2+1+1$) dynamical quarks using Wilson twisted mass fermions at maximal twist. The tuning of the strange and charm quark masses is performed at three values of the lattice spacing $a\approx 0.06$\,fm, $a\approx 0.08$\,fm and $a\approx 0.09$\,fm with lattice sizes ranging from {$L\approx 1.9$\,fm} to {$L\approx 3.9$\,fm}. We perform a preliminary study of $SU(2)$ chiral perturbation theory by combining our lattice data from these three values of the lattice spacing.}

\FullConference{The XXVIII International Symposium on Lattice Field Theory, Lattice2010\\
		June 14-19, 2010\\
		Villasimius, Italy}
 
\begin{document}

\section{Introduction}
The twisted mass formulation of Lattice QCD \cite{Frezzotti:2000nk,Frezzotti:2004wz} has been studied extensively with $N_{f}=2$ dynamical flavours by the European Twisted Mass (ETM) collaboration. In this formulation of QCD, the Wilson term is chirally rotated within an isospin doublet. The effects of the strange and charm dynamical quarks are included through a mass-split doublet as discussed in \cite{Frezzotti:2003xj,Chiarappa:2006ae,Baron:2008xa,Baron:2009zq}. Results using two of the three lattice spacings discussed in these proceedings have recently been published in \cite{Baron:2010bv}, where we describe our setup in more detail. Furthermore, at this conference, other investigations into the physics of $N_{f}=2+1+1$ twisted mass fermions have also been presented: \cite{Drach, Baron:2010vp, Urbach, Dinter, Palao}. We will briefly describe our lattice setup and recapitulate our procedure for tuning to maximal twist and the strategy for the tuning of the heavy doublet in section \ref{secaction}. We give an overview of the runs we have carried out in section \ref{secoverview}, where we also examine the status of the tuning. Finally section \ref{secresults} gives preliminary results for some observables in the light-quark sector, obtained using fits to next-to-leading order (NLO) $SU(2)$ chiral perturbation theory.
\section{Lattice setup}\label{secaction}
In the gauge sector we use the Iwasaki gauge action \cite{Iwasaki:1985we} since it improves the behavior of the lattice theory in relation to the unphysical first order phase transition for values of the hopping parameter $\kappa$ around its critical value $\kappa_{\rm crit}$ (see \cite{Baron:2010bv} and references therein). With this gauge action we observe indeed a smooth dependence of phase sensitive quantities for $\kappa\simeq\kappa_{\rm crit}$. The fermionic action for the light doublet is given by:
\begin{equation}
S_{l}=a^4\sum_x  \left\{\bar{\chi_l}(x)\left[D_W[U] + m_{0,l} + i \mu_l \gamma_5\tau_3  \right] \chi_l(x)\right\},
\end{equation}
using the notation used in \cite{Baron:2010bv}. In the heavy sector, the action becomes:
\begin{equation}\label{eqheavyact}
S_{h}=a^4\sum_{x}\left\{\bar{\chi}_{h}(x)\left[D_W[U] + m_{0,h}+i\mu_{\sigma}\gamma_{5}\tau_{1}+\mu_{\delta}\tau_{3}\right]\chi_{h}(x)\right\}.
\end{equation}
At maximal twist, physical observables are automatically $\cal{O}$$(a)$ improved without the need to determine any action or operator specific improvement coefficients. The gauge configurations are generated with a (Polynomial) Hybrid Monte Carlo updating algorithm \cite{Frezzotti:1997ym,Chiarappa:2005mx,Urbach:2005ji}, where the HMC is used for the light doublet and the PHMC for the heavy doublet.

Tuning to maximal twist requires to set $m_{0,l}$ and $m_{0,h}$ equal to some proper estimate of the critical mass $m_{\rm crit} = m_{\rm crit}(\beta)$ \cite{Frezzotti:2003xj}. As has been shown in \cite {Chiarappa:2006ae}, this is consistent with $\cal{O}$$(a)$ improvement defined by the maximal twist condition $am_{{\rm PCAC},l}=0$ (see also ref.\ \cite{Baron:2010bv}). The numerical precision at which the condition $m_{{\rm PCAC},l}=0$ is fulfilled in order to avoid residual large $\Oasq$ effects when the pion mass is decreased is, for the present range of lattice spacings, $|\epsilon/\mu_{l}| \lesssim 0.1$, where $\epsilon$ is the deviation of $m_{{\rm PCAC},l}$ from zero \cite{Boucaud:2008xu, Dimopoulos:2007qy}. As explained in \cite{Baron:2010bv}, tuning to $\kappa_{\rm crit}$ was performed independently for each set of values of $\mu_{l}$, $\mu_{\sigma}$ and $\mu_{\delta}$. From table \ref{tabruns} we observe that the estimate of $\kappa_{\rm crit}$ depends weakly on $\mu_{l}$. 
The heavy doublet mass parameters $\mu_\sigma$ and $\mu_\delta$ should be adjusted in order to reproduce the values of the renormalized $s$ and $c$ quark masses. The latter are related to $\mu_\sigma$ and $\mu_\delta$ via \cite{Frezzotti:2003xj}:
\begin{equation}
(m_{s,c})_{\rm R} = \frac{1}{\ZP} (\mu_\sigma \mp \frac{\ZP}{Z_{\rm S}} \mu_\delta),
\end{equation}
where the minus sign corresponds to the strange and the plus sign to the charm. In practice we fix the values $\mu_\sigma$ and $\mu_\delta$ by requiring the resulting kaon and $D$ meson masses to match their physical values. A detailed description of the determination of the kaon and $D$ meson masses has recently been given in \cite{Baron:2010th}.
\section{Ensemble overview}\label{secoverview}
We list in table \ref{tabruns} the action parameters for the runs considered in our current analysis. Those runs labeled with an asterisk ($*$) are ongoing at the time of the writing of these proceedings and have incomplete statistics, all other runs have around $5000$ thermalized trajectories with length $\tau=1$. Ensemble names which end in $s$ or $c$ are used to control the tuning of the strange and charm quark masses respectively. This is not an exhaustive overview of all runs performed within of our $N_f=2+1+1$ work. Other runs have been carried out in the context of tuning, reweighting, finite volume effects analysis and in order to measure the renormalization factors. 
\begin{table}[h]
\centering
\begin{tabular}{|l|cccccc|}
\hline
Ensemble & $\beta$ & $\kappa_\mathrm{crit}$ & $a\mu_l$ & $a\mu_\sigma$ & $a\mu_\delta$ & $(L/a)^3 \times T/a$\\
\hline
A30.32&1.90&0.1632720&0.0030&0.150&0.190&$32^3 \times 64$\\
A40.32&&0.1632700&0.0040&&&$32^3 \times 64$\\
A50.32&&0.1632670&0.0050&&&$32^3 \times 64$\\
A60.24&&0.1632650&0.0060&&&$24^3 \times 48$\\
A80.24&&0.1632600&0.0080&&&$24^3 \times 48$\\
A100.24&&0.1632550&0.0100&&&$24^3 \times 48$\\
A80.24s&&0.1631204&0.0080&0.150&0.197&$24^3 \times 48$\\    
A100.24s&&0.1631960&0.0100&&&$24^3 \times 48$\\
\hline
B25.32&1.95&0.1612420&0.0025&0.135&0.170&$32^3 \times 64$\\

*B35.48& &0.1612400&0.0035&&&$48^3 \times 96$\\
B35.32& &0.1612400&0.0035&&&$32^3 \times 64$\\
B55.32& &0.1612360&0.0055&&&$32^3 \times 64$\\
B75.32& &0.1612320&0.0075&&&$32^3 \times 64$\\
B85.24& &0.1612312&0.0085&&&$24^3 \times 48$\\
\hline
*D115.64&2.10& 0.1563640&0.00115&0.120&0.1385& $64^3\times128$ \\
D15.48& & 0.1563610&0.0015& & &$48^3\times96$ \\
D20.48& & 0.1563570&0.0020& & &$48^3\times96$ \\
D30.48& & 0.1563550&0.0030& & &$48^3\times96$ \\

*D45.32sc& & 0.1563550&0.0030&0.0937&0.1077&$32^3\times64$ \\
\hline
\end{tabular}
\caption{Summary of the $N_{\rm f}=2+1+1$ ensembles generated by ETMC at three values of the lattice coupling $\beta=1.90$, $\beta=1.95$ and $\beta=2.10$. From left to right, we quote the ensemble name, the value of inverse coupling $\beta$, the estimate of the critical value $\kappa_{crit}$, the light twisted mass $a\mu_l$, the heavy doublet mass parameters $a\mu_\sigma$ and $a\mu_\delta$ and the volume in units of the lattice spacing. Our notation for the ensemble names corresponds to X$\mu_l$.$L$, with X referring to the value of $\beta$ used.}
\label{tabruns}
\end{table}
\subsection{Tuning to maximal twist}
\begin{figure}[h]
\begin{centering}
\includegraphics[width=\textwidth]{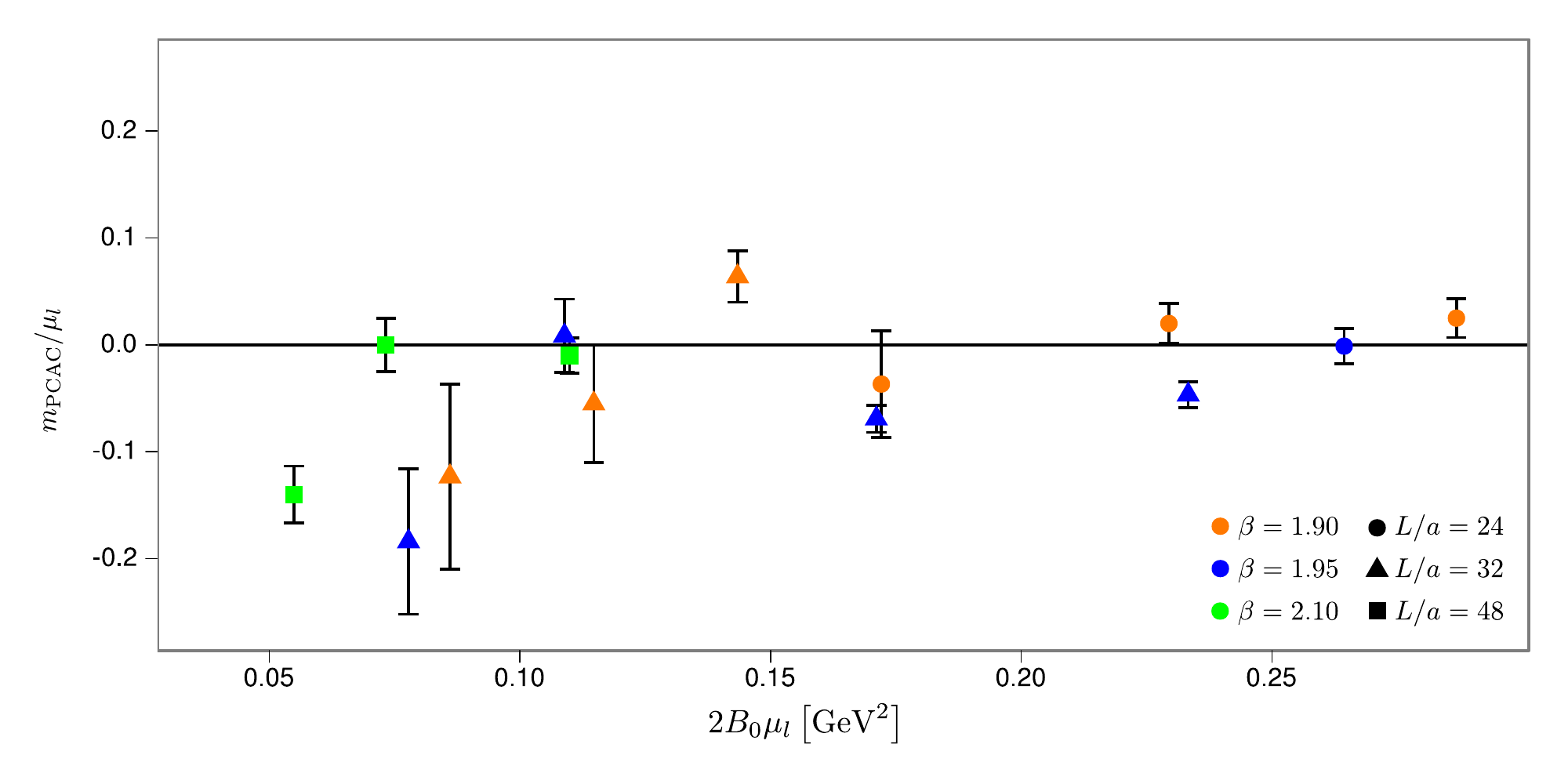} 
\caption{Status of the tuning. The ratio $m_{{\rm PCAC},l}/\mu_{l}$ is plotted as a function of the mass parameter $2B_{0}\mu_{l}$. When $|m_{{\rm PCAC},l}/\mu_{l}| \lesssim 0.1$, the ensemble is adequately tuned. Orange, blue and green symbols respectively correspond to $\beta=1.90$, $\beta=1.95$ and $\beta=2.10$ ensembles respectively.}
\label{tunfig}
\end{centering}
\end{figure}

Figure \ref{tunfig} shows the status of the tuning for the main ensembles considered in these proceedings.
\subsection{Heavy doublet tuning}
\begin{figure}[h]
\begin{minipage}[ht]{7cm}
\includegraphics[width=\textwidth]{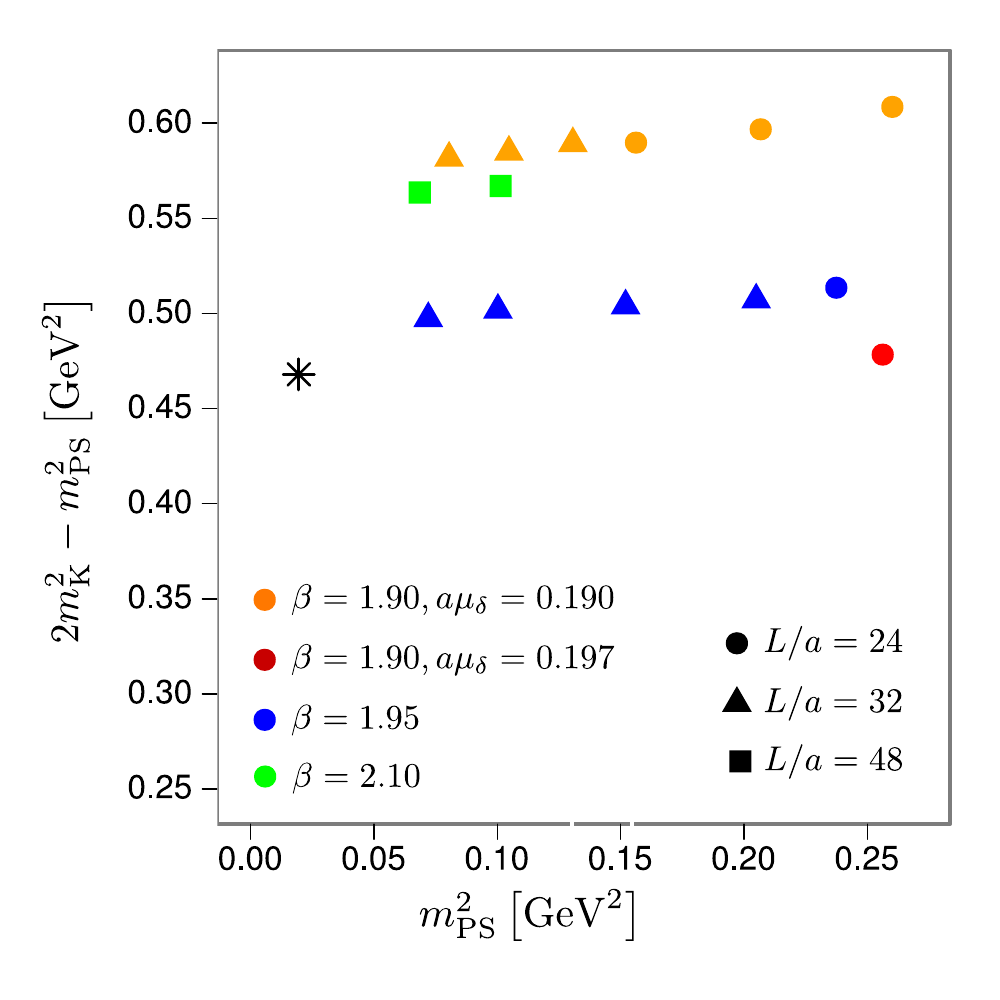} 
\end{minipage}
\begin{minipage}[ht]{7cm}
\includegraphics[width=\textwidth]{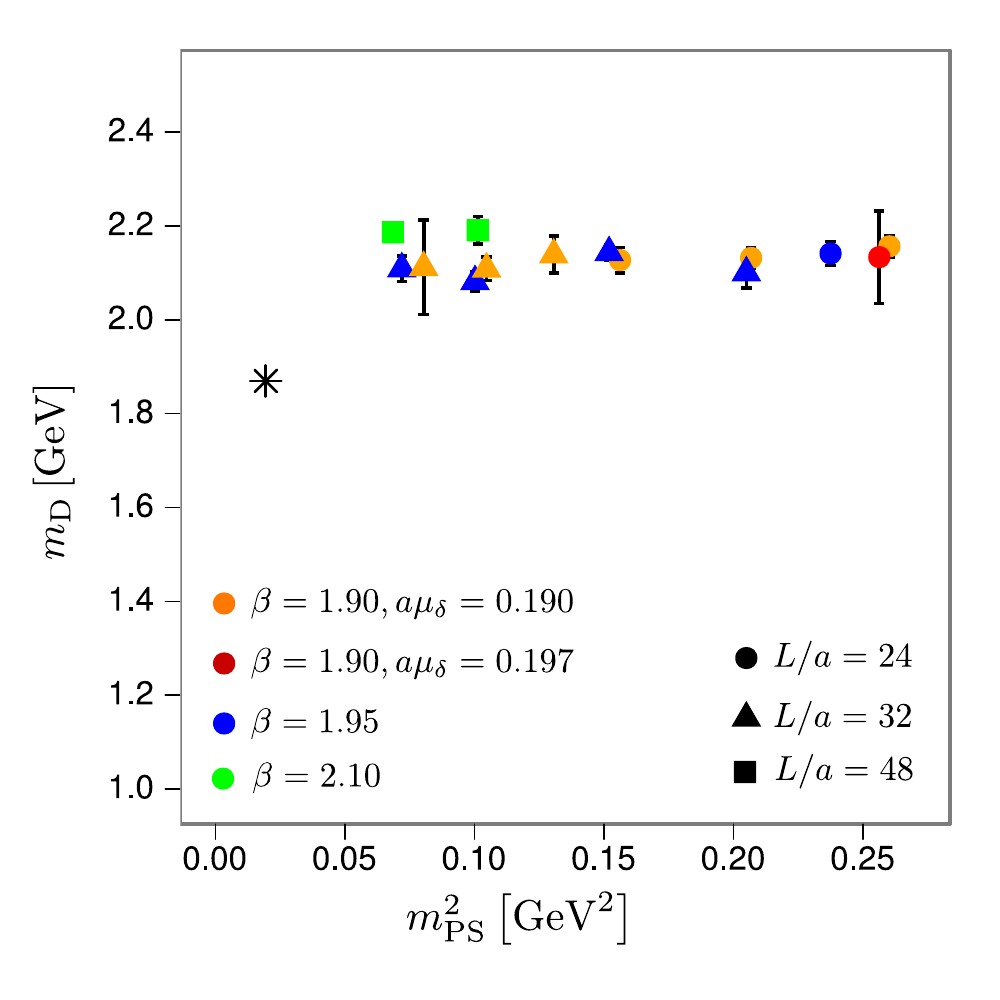} 
\end{minipage}
\caption{$2m_K^2 -\mps^2$ and $m_D$ as a function of $\mps^2$. The physical point is shown (black star) \cite{Amsler:2008zzb}. Data points have been scaled with the lattice spacing $a=0.0863(4)$\,fm for $\beta=1.90$, $a=0.0779(4)$\,fm for $\beta=1.95$ and $a=0.0607(2)$\,fm for $\beta=2.10$, where the errors quoted on the lattice spacing are only statistical.}
\label{fig:mk}\label{fig:mD}
\end{figure}

Figure \ref{fig:mk} shows the dependence of $(2 m_K^2-m_{PS}^2)$ and $m_D$ upon the light pseudoscalar mass squared for both ensembles, together with the physical point denoted by the black star. The kaon mass appears to be properly tuned at $\beta=1.95$. The
ensembles at $\beta=1.90$, $a\mu_\delta=0.190$ and $\beta=2.10$ appear to have a value of the strange quark mass larger than the physical one, while the red point at $\beta=1.90$, $a\mu_\delta=0.197$ appears to be well tuned. The $D$ meson appears heavier than in experiment for all three values of the lattice spacing. We currently have runs ongoing at both $\beta=1.90$ and $\beta=2.10$ with lower charm masses.
\section{Light meson chiral perturbation theory fits}\label{secresults}
\begin{figure}[h]
\begin{minipage}[ht]{7cm}
\includegraphics[width=\textwidth]{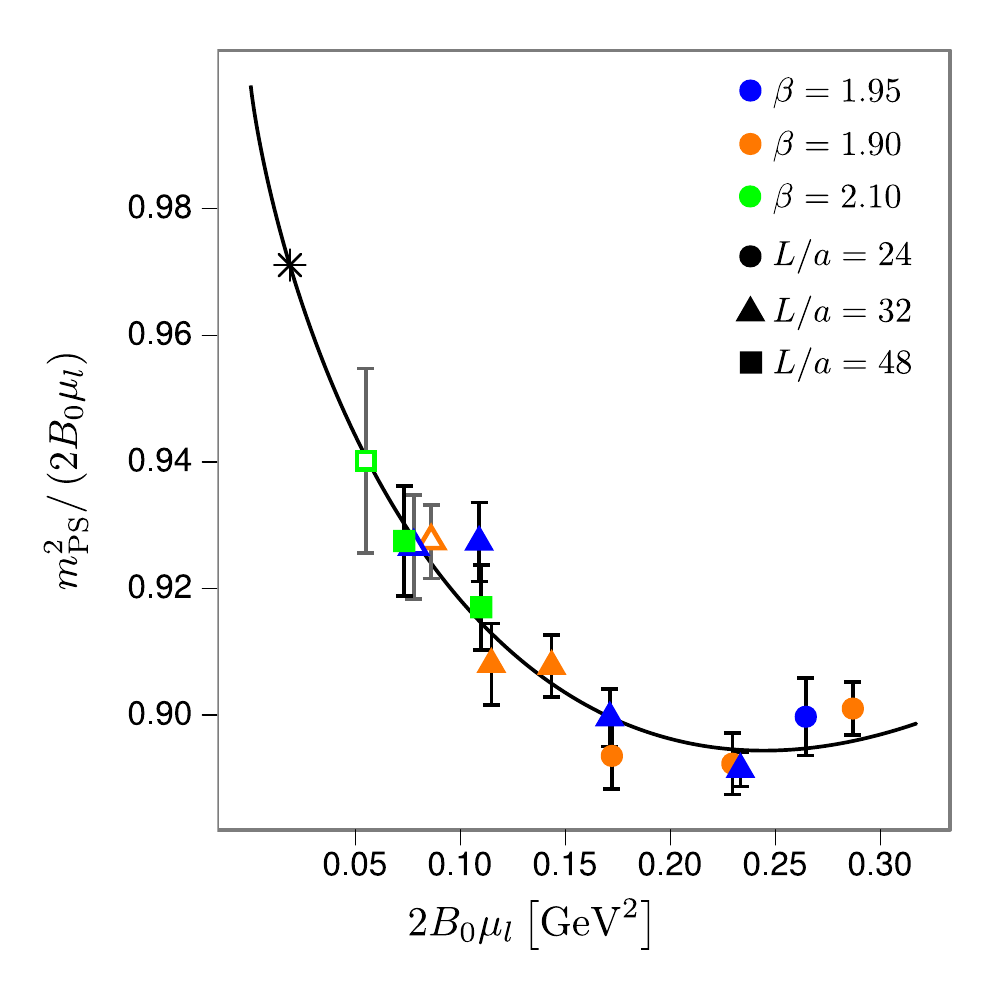} 
\end{minipage}
\begin{minipage}[ht]{7cm}
\includegraphics[width=\textwidth]{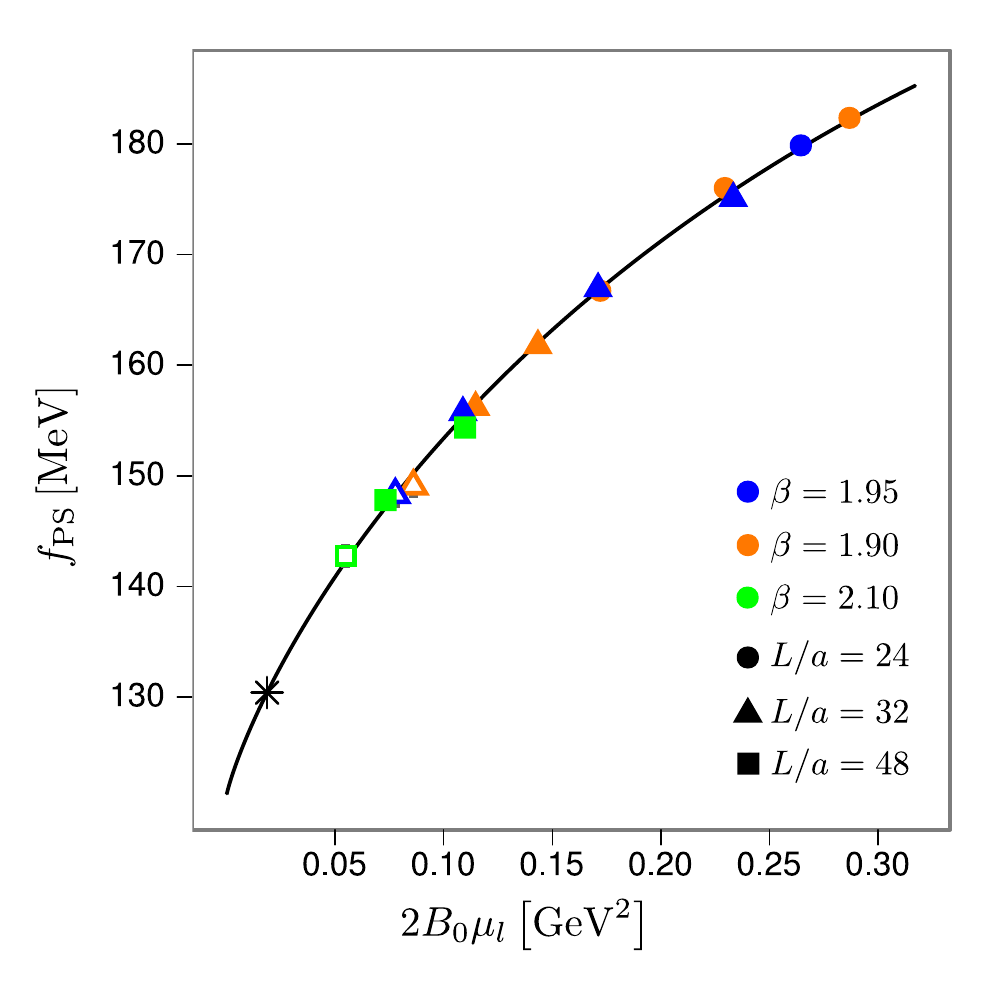} 
\end{minipage}
\caption{The charged pseudoscalar mass ratio $\mps^2 /2B_0\mu_l$ and the pseudoscalar decay constant $\fps$ as a function of the mass parameter $2B_0\mu_l$, for the combined ensembles at $\beta =1.90$, $\beta =1.95$ and $\beta =2.10$. The scale is set by $a\mu_\mathrm{phys}$, the value of $a\mu_l$ at which the ratio $\fps^{[L=\infty]}/\mps^{[L=\infty]}$ assumes its physical value \cite{Amsler:2008zzb} $f_\pi/m_\pi = 130.4(2)/135.0$ (black star). Open symbols refer to runs with full statistics, but not properly tuned to maximal twist within our criterion. Runs not at full statistics and those aimed at controlling the tuning of the strange and charm mass are not included in the plot.}
\label{mps2_B0mufig}\label{fpsfig}
\end{figure}

In order to extract the lattice spacing and light quark mass from our data, we perform a NLO $SU(2)$ chiral perturbation theory fit of the $m_{\pi}$ and $f_{\pi}$ lattice data. We group our ensembles into sets with the same lattice spacing (set $A$ at $\beta=1.90$, $B$ at $\beta=1.95$ and set $D$ at $\beta=2.10$). We have performed fits for various combinations of these sets, using the procedure described in \cite{Baron:2010bv}. We use continuum formulae and currently correct for finite volume effects as described in \cite{Colangelo:2005gd}. Note that since the quark mass enters the $\chi$PT expression, in order to combine ensembles at different lattice spacings, we need to know the renormalization factor of the quark mass $Z_{\mu}=1/\ZP$, a computation which is not yet complete. Assuming that $\ZP$ is effectively a function of $\beta$ in the range of parameters we are considering, we can fit the ratio of those $\ZP$-values and lattice spacings and extract lattice spacings from the combined fit. In every fit we use as inputs the physical values of $f_{\pi}$ and $m_{\pi}$, and extract $f_{0}$, $\bar{l}_{3}$, $\bar{l}_{4}$ and the lattice spacing. The results are listed in table \ref{tabchipt}, while figure \ref{mps2_B0mufig} shows the fit to sets $A$, $B$ and $D$ combined.

\begin{table}[h]
\begin{center}
\begin{tabular}{|c|c|c|c|c|c|c|c|}
\hline
set & pts & $f_{0}$(MeV) & $\bar{l}_{3}$ & $\bar{l}_{4}$ & $a_{\beta=1.90}$(fm) & $a_{\beta=1.95}$(fm) & $a_{\beta=2.10}$(fm)\\
\hline
$A$ & $5$ & $120.96(7)$ & $3.44(6)$ & $4.77(2)$ & $0.0859(5)$ & & \\
$B$ & $4$ & $121.15(8)$ & $3.70(7)$ & $4.67(3)$ & & $0.0782(6)$ & \\
$A$ \& $B$ & $9$ & $121.03(5)$ & $3.54(5)$ & $4.74(2)$ & $0.0861(4)$ & $0.0778(4)$ & \\
$A$ \& $D$ & $7$ & $120.99(7)$ & $3.42(7)$ & $4.76(3)$ & $0.0861(5)$ & & $0.0606(3)$\\
$B$ \& $D$ & $6$ & $121.20(8)$ & $3.68(7)$ & $4.65(3)$ & & $0.0785(6)$ & $0.0609(3)$\\
$A$ \& $B$ \& $D$ & $11$ & $121.05(5)$ & $3.53(5)$ & $4.73(2)$ & $0.0863(4)$ & $0.0779(4)$ & $0.0607(2)$\\
\hline
\end{tabular}
\end{center}
\caption{Results from the NLO $SU(2)$ $\chi$PT fits for various combinations of the ensembles. Errors are statistical only, extracted from 200 bootstrap samples. The column "pts" refers to the number of ensembles used in that fit.}
\label{tabchipt}
\end{table}

The observed agreement between the extracted parameters suggests that our data for $\mps$ and $\fps$ are fairly well described by NLO $SU(2)$ chiral perturbation theory. Using the spread of parameters as a rough estimate of the systematic error, it appears to be smaller than the statistical error for all quantities, with the exception of $\bar{l}_{3}$. A more complete analysis of the systematic effects (analogous to \cite{Herdoiza09}) is in progress. We expect to extend our analysis by including twisted mass chiral perturbation theory formulae as described in \cite{Bar:2010jk} and to use twisted mass finite volume effects formulae \cite{Colangelo:2010cu} when our neutral pion measurements are more complete.

\section{Summary and Outlook}
We have given an update of the status of the runs performed by the ETM Collaboration using $N_f=2+1+1$ flavours of Wilson twisted mass fermions. We have given first results at a new finer ($\beta=2.10$) lattice spacing and attempted to combine them with existing datasets at two other lattice spacings ($\beta=1.90$ and $\beta=1.95$). The production of ensembles at the finest lattice spacing is still ongoing. As already stated, a complete control of the different systematic effects present in chiral fits of pion observables is still missing. ETMC is currently pursuing the direct determination of the renormalisation factor $\ZP$ \cite{Palao} appearing in the fits combining ensembles at several values of the lattice spacing. The preliminary results presented in this work are nevertheless very encouraging and suggest a fairly good description of our lattice data for $\mps$ and $\fps$ by NLO $SU(2)$ chiral perturbation theory.

\acknowledgments
We thank the members of the ETM Collaboration for valuable discussions. The HPC resources for this project have been made available by the computer centres of Barcelona, Groningen, J\"ulich, Lyon, Munich, Paris and Rome (apeNEXT), which we thank for enabling us to perform this work. This work has also been supported in part by the DFG Sonderforschungsbereich/Transregio SFB/TR9-03, and by GENCI (IDRIS - CINES), Grant 2009-052271.

\end{document}